\documentclass{PoS}
\usepackage{amsmath,amssymb,amsfonts,macros,patricks}

\title{B physics from HQET in two-flavour lattice QCD}
\ShortTitle{$\Nf=2$ Lattice HQET}

\author{\includegraphics[width=2.5cm,angle=0]{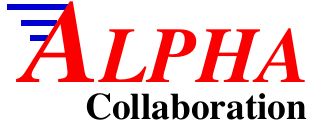}\hfill\parbox{3.5cm}{\footnotesize\it CERN-PH-TH/2012-289\\ DESY 12-188\\ HU-EP-12/35 \\
LPT Orsay/12-107 \\ MS-TP-12-14 \\ SFB/CPP-12-78 \\ TCDMATH 12-09
}}
\author{\speaker{F.~Bernardoni}$\,^{a}$, B.~Blossier$\,^{b}$,
J.~Bulava$\,^{c}$, M.~Della\ Morte$\,^{d}$, P.~Fritzsch$\,^{e}$,
N.~Garron$\,^{f}$, A.~G\'erardin$\,^{b}$, J.~Heitger$\,^{g}$,
G.~von~Hippel$\,^{d}$, H.~Simma$\,^{a}$, R.~Sommer$\,^{a}$\\
$^a$~NIC, DESY, Platanenallee~6, 15738~Zeuthen, Germany\\
$^b$~Laboratoire~de~Physique~Th\'eorique, CNRS/Universit\'e~Paris~XI,
F-91405~Orsay~Cedex, France\\
$^c$~CERN, Physics~Department, TH~Division, CH-1211~Geneva~23, Switzerland\\
$^d$~Institut~f\"ur~Kernphysik, University~of~Mainz, Becher-Weg~45, 55099~Mainz, Germany\\
$^e$~Institut~f\"ur~Physik, Humboldt-Universit\"at~zu~Berlin, Newtonstr.~15, 12489~Berlin, Germany\\
$^f$~School~of~Mathematics, Trinity~College, Dublin~2, Ireland\\
$^g$~Universit\"at~M\"unster, Institut~f\"ur~Theoretische~Physik, Wilhelm-Klemm-Str.~9, 48149~M\"unster, Germany
\\
\email{fabio.bernardoni@desy.de}}

\abstract{We present our analysis of B physics quantities using non-perturbatively
          matched Heavy Quark Effective Theory (HQET) in $\Nf=2$ lattice QCD
          on the CLS ensembles. Using all-to-all propagators, HYP-smeared
          static quarks, and the Generalized Eigenvalue Problem (GEVP) approach
          with a conservative plateau selection procedure,
          we are able to systematically control all sources of error.
          With significantly increased statistics compared to last year,
          our preliminary results are
          $\overline{m}_{\rm b}(\overline{m}_{\rm b}) = 4.22(10)(4)_{z}\,\GeV$ 
          for the $\MSbar$ b-quark mass, and
          $\fB =  193(9)_{\rm stat}(4)_{\chi}\, \MeV$ and
          $\fBq{s} =  219(12)_{\rm stat}\, \MeV$ for the B-meson decay constants.
          \vskip1cm\hfill\includegraphics[width=2cm,keepaspectratio=]{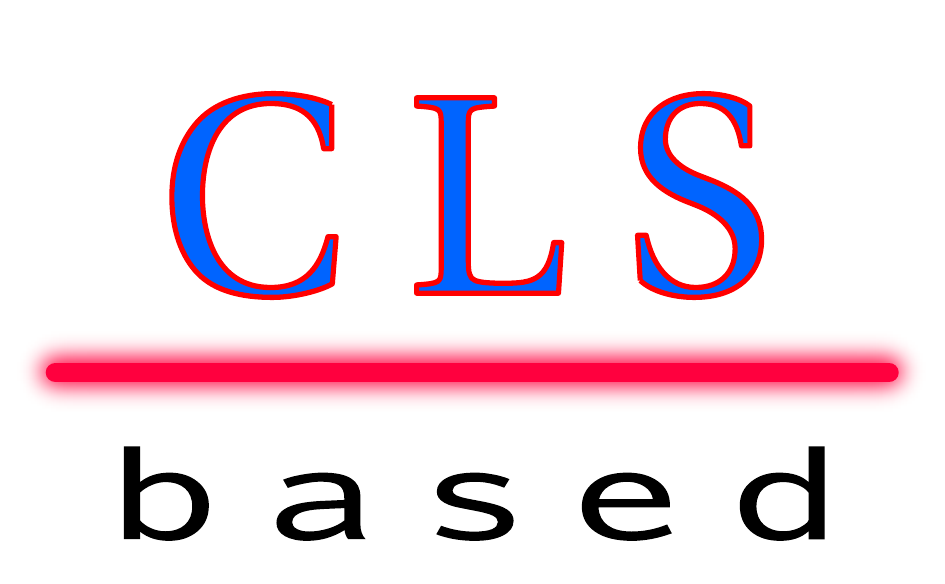}}

\FullConference{The 30th International Symposium on Lattice Field Theory - Lattice 2012,\\
		June 24-29, 2012\\
		Cairns, Australia}

\begin{document}


\section{Introduction}

Weak decays of heavy mesons constrain the CKM 
matrix encoding the flavour-changing weak interactions.
Besides experimental data, lattice QCD results for
low-energy hadronic matrix elements decisively contribute to
precision tests in the beauty sector. Since the significance of these tests
is predominantly limited by theoretical uncertainties, 
lattice computations with an overall accuracy of a few percent are
highly desirable.

\begin{figure}
\begin{center}
\includegraphics[width=0.5\textwidth,keepaspectratio=]{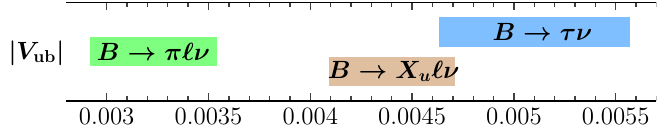}
\end{center}
\caption{
  An illustration of the observed tension between the
  determinations of $|V_{\rm ub}|$ from leptonic, exclusive semileptonic,
  and inclusive semileptonic channels; $\pm 1\sigma$ bands are shown.}
\end{figure}

A quantity that has attracted much attention in this context recently
is the CKM matrix element $|V_{\rm ub}|$. It can be determined in different
ways, in particular from 
inclusive semi-leptonic processes ${B\to X_{\rm u}\ell\nu}$
(whose theoretical treatment involves the use of the optical theorem,
the heavy quark expansion and perturbation theory),
from exclusive semi-leptonic decays ${B\to\pi\ell\nu}$
involving the hadronic form factor $f_{+}(q^2)$,
and from exclusive leptonic decays ${B\to\tau\nu}$
involving the hadronic decay constant $\Fb$
(where the latter two cases require lattice determinations of $f_{+}(q^2)$
and $\Fb$ in order to extract $|V_{\rm ub}|$).
Currently, there is a $\sim 3\sigma$ tension between the two exclusive 
(semi-leptonic and leptonic) determinations of $|V_{\rm ub}|$, as well as
tension with the estimate from inclusive B-meson decays.
Precise and reliable lattice calculations with good control of systematic
errors are required in order to answer the question whether this tension
really hints at New Physics in the B-sector.

In order to describe B physics on the lattice with controlled errors,
it is necessary to cover multiple physical scales differing by several
orders of magnitude; in principle it would be desirable to have
$\;\Lambda_{\rm IR}\,=\,L^{-1}\;\ll\; \mpi\,,\,\ldots\,,\,\mB\;\ll\;
a^{-1}\,=\,\Lambda_{\rm UV}$.
What is achievable in practice is $L\gtrsim 4/\mpi\approx 6\,\Fm$
to suppress finite-size effects for light quarks, and
$a\lesssim 1/(2\mD)\approx 0.05\,\Fm$ to tame discretization 
errors in the charm sector and allow the use of relativic c-quarks.
However, the b-quark scale $\mb\sim 4\mc$ has to be separated from the
others in a theoretically sound way before simulating the theory;
in this work, we employ the Heavy Quark Effective Theory (HQET)
formulation~\cite{HQET} for the b-quark in heavy-light systems.

In this update on our earlier analysis~\cite{lat11:patrick}, we are now able to
include more statistic, as well as new measurements including a strange
valence quark (although not all of the latter are fully analysed yet).
The newest analysis also benefits from an improved estimate of the matching
scale $L_1=0.401(13)\Fm$, which previously dominated our eror on $\mb$,
as well as a refinement of the procedure used to select the plateau regions
from which we extract our estimates of the large-volume observables.

\section{Methods}

\subsection{Non-perturbative HQET at $\Or(1/\mb)$}

By performing a systematic asymptotic expansion of the QCD Lagrangian
in $\lQCD/\mb\ll 1$, and truncating $\Or\big(\lQCD^2/\mb^2\big)$ contributions,
one arrives at the (continuum) HQET Lagrangian:
\be
\lag{HQET}(x)
=
\;\displaystyle\heavyb(x)\,D_0\,\heavy(x)
\;-\,{\omkin}\Okin (x) -{\omspin}\Ospin(x)\,,
\ee\be
\Okin(x)  = \heavyb(x)\,\vecD^2\,\heavy(x)\,,
~~~~~~~ \Ospin(x) = \heavyb(x)\,\vecsigma\cdot\vecB\,\heavy(x).\nonumber
\ee
Similarly, the time component of the heavy-light axial current $A_0$
(at ${\bf p}=0$) expands as
\bea
\Arenhqet= 
{\zahqet}\,\Big[\,\Astat+{\cah{1}}\Ah{1}\,\Big]\;\;,\;\; 
\Astat=
\lightb\,\gamma_0\gamma_5\,\heavy\;\;,\;\;
\Ah{1}= 
\lightb\,\gamma_5\gamma_i\,\half\,(\nabsym{i}\,-\!\lnabsym{i}\,)\,\heavy.
\eea
In the HQET approach, the $\minv$--terms appear as local operator insertions
in correlation  functions, and therefore the renormalizability of the static
theory carries over to HQET (at each order in $\minv$). This ensures the
existence of the continuum limit, once the HQET parameters
$\,{\omega_i}\in\big\{{ m_{\rm bare}}\,,{\zahqet},
{\cah{1}},{\omkin},{\omspin}\big\}$
have been fixed through \emph{non-perturbative} matching~\cite{HQET:pap1}
so that no uncancelled power divergences in $a^{-1}$ (which are induced
by operator  mixing in the effective theory) remain that would spoil taking
the continuum limit.

The strategy of non-perturbative matching \cite{HQET:param1m,HQET:Nf2param1m}
is as follows: The matching is performed in the Schr\"odinger Functional scheme
in a small volume $L_1\approx 0.4\,\Fm$, where due to $a\mb\ll 1$ simulations
with a relativistic b-quark are feasible. The HQET parameters $\omega_i$ are
fixed by imposing the matching conditions
\bea
\PhiHQET(z,a)=\PhiQCD(z,0)\,,~~~~~
\PhiQCD(z,0)=\lim_{a\to 0}\PhiQCD(z,a)\,,
\eea
so that $\omega_i$ inherit the quark mass dependence from non-perturbatively
renormalized QCD via their dependence on $z\equiv L_1M$, where $M$ is the RGI
quark mass~\cite{impr:babp_nf2}.
We then use a recursive finite-size scaling procedure to perform the step
$L_1\to L_2=2L_1$ and finally to make contact to physically large volumes
$L_{\infty}\gtrsim \max(2~\Fm,4/m_\pi)$.

As a result of this procedure~\cite{HQET:Nf2param1m}, the
$\nf=2$ HQET parameters $\omega_i(z,a)$ (which absorb the power divergences
of HQET) are non-perturbatively known for a number of $z$--values around the
b mass at the lattice spacings used in our large--volume simulations.

\subsection{Large volume computations and techniques}

Our large--volume measurements are performed on the $\nf=2$ CLS ensembles,
which use the plaquette gauge action and non-perturbatively $\Or(a)$ 
improved Wilson quarks and were generated using the
DD-HMC~\cite{ddhmc+deflat:luescher} and/or the MP-HMC~\cite{mphmc:hasenb1+marina}
algorithms. The ensembles fulfill the condition $L\mpi\gtrsim 4$,
and we use a range of pion masses $(190\lesssim\mpi\lesssim 440)\,\MeV$
at three lattice spacings $(0.05\lesssim a\lesssim 0.08)\,\Fm$,
where the scale has been set through $f_{\rm K}$~\cite{scale:fK_Nf2}.

In the computation of the static-light correlation functions, we use
HYP smearing for the static quarks~\cite{HYP:HK01+ALPHA}
as well as a variant of the stochastic all-to-all propagator method for the
relativistic quarks
(with multiple noise sources per configuration
and full time-dilution)~\cite{ata:dublin,HQET:msplit+fb1m}
in order to improve statistical precision.

\begin{table}
\begin{center}
\begin{center}
\renewcommand{\arraystretch}{1.25}\small
\begin{tabular}{|c|c|c|c|c|c|}
\hline
$\beta$&$a\,[\,\Fm\,]$ &$L^{3}\times T$&$m_{\pi}\,[\,\MeV\,]$&\#\\
\hline
\hline
5.2&0.075&$32^{3}\times 64$&380&1000\\
&&$32^{3}\times 64$&330&500\\
\hline
5.3&0.065&$32^{3}\times 64$&440&1000\\
&&$48^{3}\times 96$&310&500\\
&&$48^{3}\times 96$&270&600\\
&&$\bf 64^{3}\times 128$&\bf 190&\bf 600\\
\hline
5.5&0.048&$48^{3}\times 96$&440&400\\
&&$\bf 48^{3}\times 96$&\bf 340&\bf 900\\
&&$64^{3}\times 128$&270&900\\
\hline
\end{tabular}
\end{center}
\end{center}
\caption{Overview of the CLS configurations used in this analysis. The
         ensembles shown in bold are new to the present analysis.}
\end{table}

To control excited state contaminations to the HQET energies and
matrix elements, we solve the Generalized Eigenvalue Problem
(GEVP)~\cite{HQET:gevp,HQET:msplit+fb1m} 
\begin{align}
  C(t)v_n(t,t_0) = \lambda_n(t,t_0)C(t_0)v_n(t,t_0)\,, ~~~~~~~~~ t_0<t<2t_0\,,
\end{align}
for an $N\times N$ correlator matrix $C(t)$ with $N=3$, and derive the energies
and matrix elements from the eigenvalues and eigenvectors $\lambda_n$, $v_n$,
such that the corrections to the energies behave like
$\propto \exp\left\{ -(E_{N+1}-E_1)t\right\}$ and the corrections to the
matrix elements like 
$\propto \exp\left\{ -(E_{N+1}-E_1)t_0\right\}
\times\exp\left\{ -(E_2-E_1)(t-t_0)\right\}$;
for details see~\cite{HQET:gevp,lat11:patrick}.
We minimize our systematic errors by a conservative choice of plateau ranges:
for fixed $t_{\rm max}$ we vary $t_{\rm min}$ such that the errors
$\sigma(t_{\rm min})$ of the plateau average $A(t_{\rm min})$
fulfil $\sigma_{\rm stat}\gtrsim 3\sigma_{\rm sys}$.
Specifically, for each value of $t_{\rm min}$ we compute
\begin{equation}
r(t_{\rm min})=\frac{|A(t_{\rm min})-A(t_{\rm min}-\delta)|}{\sqrt{\sigma^2(t_{\rm min})+\sigma^2(t_{\rm min}-\delta)}}\,,
\end{equation}
where $\delta=\frac{2}{3}r_0\approx 2/(E_{N+1}-E_1)$ is chosen such that
we expect the influence of excited-state contributions to have decayed
by a factor $\sim \rme^2$, and take the first value of $t_{\rm min}$
satisfying $r(t_{\rm min})\le 3$.

\begin{figure}
\includegraphics[width=0.99\textwidth,keepaspectratio=]{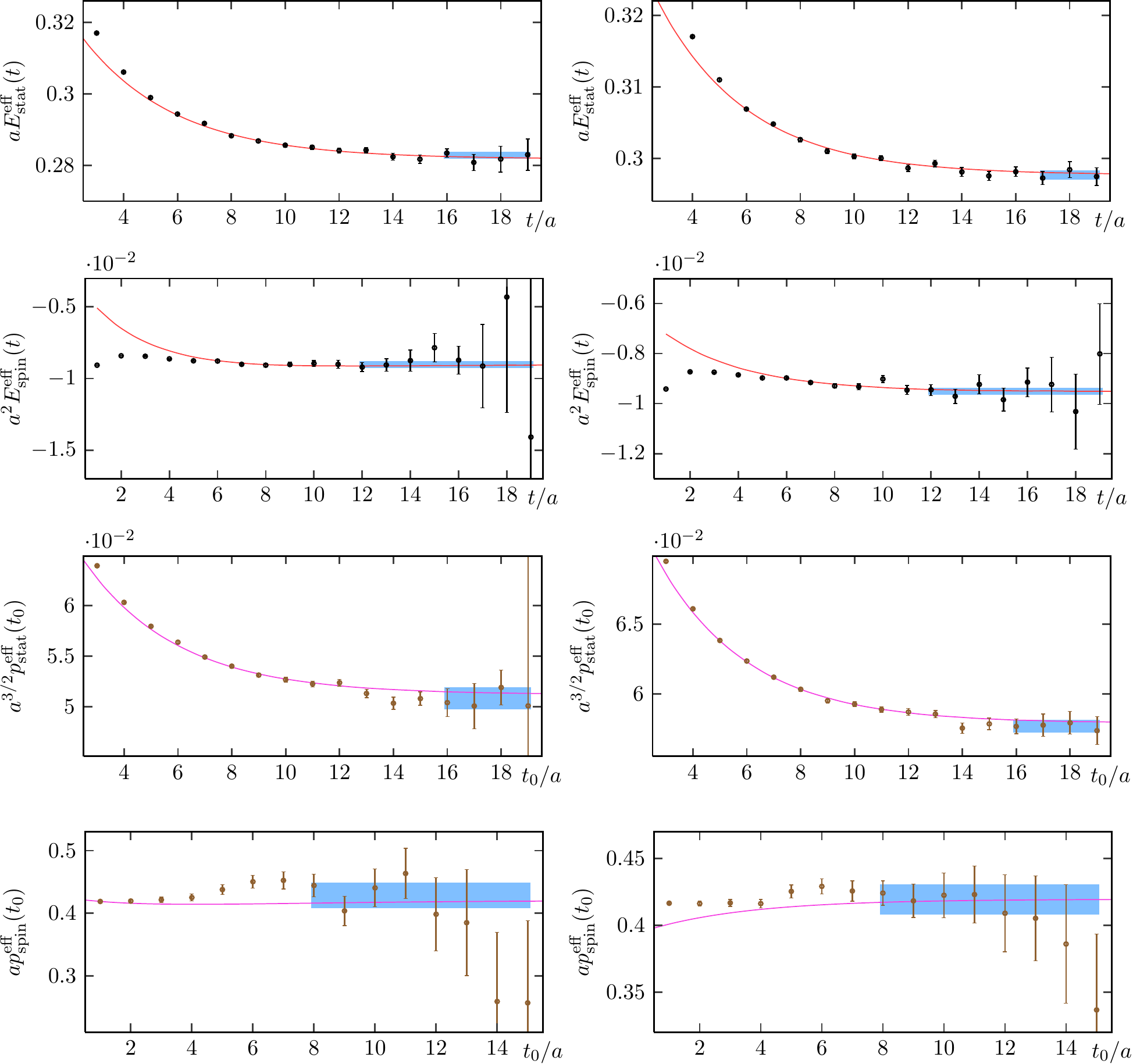}
\caption{Examples of preliminary results from an ensemble with 
         $a=0.048\,\Fm$, $\mpi\simeq 340\,\MeV$ and
         $L^{3}\times T=48^{3}\times 96$;
         {\em left:} HYP2 static-light; {\em right:} HYP2 static-strange.
         Shown are the effective masses or matrix elements as a function of
         $t=t_0+a$, together with a band indicating the extracted plateaux.
         The result of a global fit to the data at $t_0/a\ge 6$, which
         also includes a large number of data points not shown here
         (including results from $t>t_0+a$, and from $N=4,5$),
         is also displayed as a solid curve, showing good agreement between
         the fit and the more conservative analysis ultimately employed.}
\end{figure}

\section{Results}

Combining the HQET parameters with the GEVP results for matrix elements and 
energies, we obtain observables depending on the pseudoscalar (sea) mass
$m_\pi$ and lattice spacing $a$, as well as (through the HQET parameters)
on the heavy quark mass parameter $z$.

\subsection{Mass of the b-quark}
We fix $m_{\rm b}$ by imposing
  $\mB(z_{\rm b},m_{\pi}^{\rm exp},a=0) \equiv \mB^{\rm exp} = 5279.5\MeV$
through the fit ansatz
\begin{align}
 \mB\left(z,m_\pi,a,{\rm HYPn}\right) &= B(z)+ C m^2_\pi - \frac{3\widehat{g}^2}{16\pi f_\pi^2} m^3_\pi  + D_{\rm HYPn} a^2 ,
&  \widehat{g} =0.51(2)\,\cite{Bulava:2010ej}.
\end{align}
Using the NLO mass definition of HQET,
$\mB ={ m_{\rm bare}} + E^{\rm stat} + { \omega_{\rm kin}}  E^{\rm kin}+ { \omega_{\rm spin}} E^{\rm spin}$, we find
\begin{align}
   z_{\rm b} &= 13.34(33)(13)_{z}\;, & &\text{or equivalently} &
   \overline{m}_{\rm b}(\overline{m}_{\rm b}) &= 4.22(10)(4)_{z}\,\GeV  \;. 
\end{align}
Having fixed the physical mass of the b-quark, we interpolate the HQET parameters to $z\equiv z_{\rm b}$.

\subsection{B-meson decay constants at NLO of HQET}

\begin{figure}
\begin{center}
\includegraphics[width=\textwidth,keepaspectratio=]{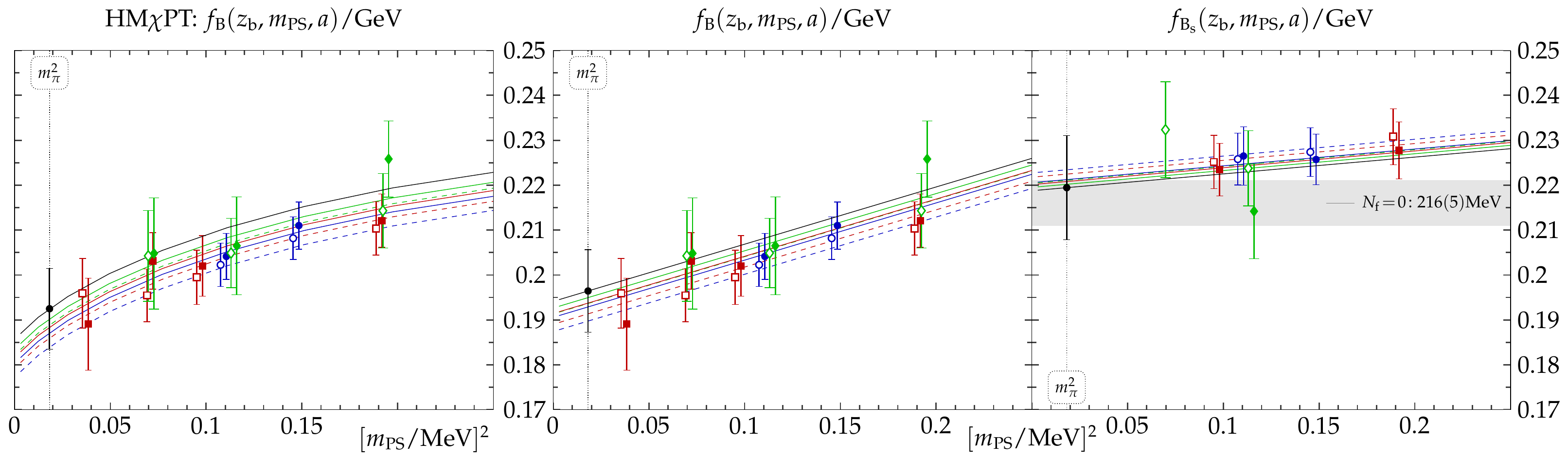}
\end{center}
\caption{{\em Left:} HM$\chi$PT extrapolation of $\fB$;
         {\em centre/right:} linear extrapolation of $\fB$ and $\fBs$.
         The blue, red and green points correspond to ensembles at
         $a=0.075$~fm, $0.065$~fm and $0.048$~fm, respectively.
         Filled symbols denote HYP2, empty symbols HYP1;
         the joint continuum and chiral extrapolation is shown in black,
         with the fit formulae evaluated at each given lattice spacing
         shown in colour (solid for HYP2, dashed for HYP1).}
\end{figure}

We determine the ${\rm B}_{\rm r}$-meson decay constant for light (${\rm r}=d$) and strange (${\rm r}=s$) quarks through
\begin{align}
        \ln(a^{3/2}\fBq{r}\sqrt{\mBq{r}/2}) &=  { \ln(\zahqet)} + \ln(a^{3/2}p^{\rm stat}_{\rm r}) + b^{{\rm stat}}_{\rm A} am_{\rm q,r}
        + { \omega_{\rm kin}}  p^{\rm kin}_{\rm r} + { \omega_{\rm spin}}  p^{\rm spin}_{\rm r} + { \cah{1}} p^{\rm A^{(1)}}_{\rm r} \,.
\end{align}
In order to estimate a systematic error in our combined chiral and continuum extrapolation, we use both a fit motivated by Heavy Meson Chiral Perturbation Theory (HM$\chi$PT)
\cite{HMxPT:G92,HMxPT:SZ96}
and a linear fit in $m_{\pi}^2$:
\begin{align}
        \fBq{r}\left(m_{\pi},a,{\rm HYPn}\right) &= b_{\rm r}+c_{\rm r}\, m^2_{\pi} + d_{\rm r,HYPn}\, a^2 \,,  & (\mbox{linear}) \\
 \fB\left(m_{\pi},a,{\rm HYPn}\right) &= b'\left[1-\frac{3}{4} \frac{1+3\widehat{g}^2}{(4\pi f_\pi)^2}
          m^2_{\pi}\ln (m^2_{\pi})\right] + c' m^2_{\pi}   + d_{\rm HYPn}' a^2 \,.  & (\mbox{HM}\chi\mbox{PT})
\end{align}
Both of these fit formulae are based on a simultaneous expansion
in $a$ and $1/\mb$, where $\rmO(a)$ discretization effects are dropped, since
they are also $\rmO(1/\mb)$; for $\fB$, only the non-analytic terms coming
from the lowest order in $1/\mb$ are retained.

Our analysis for Lattice~2012 (where for $\fBq{s}$ not all ensembles are analysed yet) gives
\bean
     \fB =  193(9)_{\rm stat}(4)_{\chi}\, \MeV\,, ~~~~~~~
 \fBq{s} =  219(12)_{\rm stat}\, \MeV .
\eean
The quenched value was $\fBq{s}=216(5)$ (using $\r_0=0.5\fm$), indicating
small quenching effects.
Note that our present $\Nf=2$ estimate of $\fB$ is about one standard deviation
larger than the previous value from~\cite{lat11:patrick}, due both to increased
statistics and the inclusion of additional sea quark masses. A further
improvement is an improved control of systematic errors from excited-state
contributions (which enlarges the statistical errors).

\section{Conclusions}

We have presented results with a significant increase in statistics and
improved control of systematic errors. We obtain a value for $\fB$ that
is similar to those obtained by other collaborations
\cite{Na:2012kp,Bazavov:2011aa,McNeile:2011ng,Dimopoulos:2011gx}.
In the case of $\fBq{s}$, we find quenching effects to be undetectable,
while for $\fB$ such a comparison is not meaningful due to the pathological
chiral behaviour of the quenched theory.

We are now finalizing our analysis, and more detailed publications are
forthcoming.
We are also investigating further spectral quantities within our approach,
such as the B-meson spin splittings, and are preparing for the determination
of $B\to$ light semileptonic form factors
\cite{lat12:fabio}.
%


\section*{Acknowledgments}

\small

This work is supported in part
by the SFB/TR~9
and grant HE~4517/2-1 (P.F. and J.H.)
of the Deutsche Forschungsgemeinschaft
and by the European Community through
EU Contract MRTN-CT-2006-035482, ``FLAVIAnet''.
P.F. thanks the ECT* in Trento for support during the workshop
``Beautiful Mesons and Baryons on the Lattice''.
We thank our colleagues in the CLS effort for the joint production
and use of gauge configurations.
We gratefully acknowledge the computer resources provided
within the Distributed European Computing Initiative
by the PRACE-2IP,
with funding from the European Community's Seventh Framework Programme
(FP7/2007-2013) under grant agreement RI-283493,
by the Grand \'Equipement National de Calcul Intensif at CINES in Montpellier,
and by the John von Neumann Institute for Computing
at FZ~J\"ulich, at the HLRN in Berlin, and at DESY, Zeuthen.




\small


\begin{thebibliography}{10}

\bibitem{HQET}
E. Eichten and B. Hill,
\newblock Phys. Lett. B234 (1990) 511;
E. Eichten and B. Hill,
\newblock Phys. Lett. B 243 (1990) 427.

\bibitem{lat11:patrick}
B. Blossier {\em et al.},
\newblock PoS LATTICE2011 (2011) 280, arXiv:1112.6175.

\bibitem{HQET:pap1}
J. Heitger and R. Sommer,
\newblock JHEP 0402 (2004) 022, hep-lat/0310035.

\bibitem{HQET:param1m}
B. Blossier {\em et al.},
\newblock JHEP 1006 (2010) 002, arXiv:1001.4783.

\bibitem{HQET:Nf2param1m}
B. Blossier {\em et al.},
\newblock JHEP 1209 (2012) 132, arXiv:1203.6516.

\bibitem{impr:babp_nf2}
P. Fritzsch, J. Heitger and N. Tantalo,
\newblock JHEP 1008 (2010) 074, arXiv:1004.3978.

\bibitem{ddhmc+deflat:luescher}
M. {L\"uscher},
\newblock Comput. Phys. Commun. 156 (2004) 209, hep-lat/0310048;
\newblock Comput. Phys. Commun. 165 (2005) 199, hep-lat/0409106;
\newblock JHEP 0712 (2007) 011, arXiv:0710.5417.

\bibitem{mphmc:hasenb1+marina}
M. Hasenbusch,
\newblock Phys. Lett. B 519 (2001) 177, hep-lat/0107019;
M. Marinkovic and S. Schaefer,
\newblock PoS LATTICE2010 (2010) 031, arXiv:1011.0911.

\bibitem{scale:fK_Nf2}	
P. Fritzsch {\em et al.},
\newblock Nucl.Phys. B865 (2012) 397, arXiv:1205.5380.

\bibitem{ata:dublin}
J. Foley {\em et al.},
\newblock Comput. Phys. Commun. 172 (2005) 145, hep-lat/0505023.

\bibitem{HQET:gevp}
B. Blossier {\em et al.},
\newblock JHEP 0904 (2009) 094, arXiv:0902.1265.

\bibitem{HQET:msplit+fb1m}
B. Blossier {\em et al.},
\newblock JHEP 1005 (2010) 074, arXiv:1004.2661;
B. Blossier {\em et al.},
\newblock JHEP 1012 (2010) 039, arXiv:1006.5816.

\bibitem{HYP:HK01+ALPHA}
A. Hasenfratz and F. Knechtli,
\newblock Phys. Rev. D 64 (2001) 034504, hep-lat/0103029;
M. Della Morte {\em et al.},
\newblock Phys. Lett. B 581 (2004) 93, hep-lat/0307021;
M. Della Morte, A. Shindler and R. Sommer,
\newblock JHEP 0508 (2005) 051, hep-lat/0506008.

\bibitem{HMxPT:G92}
J.L. Goity,
\newblock Phys. Rev. D46 (1992) 3929, hep-ph/9206230.

\bibitem{HMxPT:SZ96}
S.R. Sharpe and Y. Zhang,
\newblock Phys. Rev. D53 (1996) 5125, hep-lat/9510037.

\bibitem{Bulava:2010ej}
J. Bulava, M.A. Donnellan and R. Sommer,
\newblock PoS LATTICE2010 (2010) 303, arXiv:1011.4393.

\bibitem{Na:2012kp}
H. Na {\em et al.} [HPQCD Collaboration],
\newblock Phys.\ Rev.\ D {\bf 86} (2012) 034506, arXiv:1202.4914.

\bibitem{Bazavov:2011aa}
A. Bazavov {\it et al.}  [Fermilab Lattice and MILC Collaboration],
\newblock Phys.\ Rev.\ D {\bf 85} (2012) 114506, arXiv:1112.3051.

\bibitem{McNeile:2011ng}
C. McNeile {\em et al.} [HPQCD collaboration],
\newblock  Phys.\ Rev.\ D {\bf 85} (2012) 031503, arXiv:1110.4510.

\bibitem{Dimopoulos:2011gx}
P. Dimopoulos {\it et al.}  [ETM Collaboration],
\newblock JHEP {\bf 1201} (2012) 046, arXiv:1107.1441.

\bibitem{lat12:fabio}
F. Bernardoni {\em et al.} [ALPHA Collaboration],
\newblock in these proceedings, arXiv:1210.3478.

\end{thebibliography}
\end{document}